\documentclass[intlimits,twoside,a4paper]{article}

\usepackage[cp1251]{inputenc}
\usepackage[eqsecnum]{cmpj3}
\usepackage{bm}

\issue{2020}{23}{3}{33707}
\doinumber{10.5488/CMP.23.33707}
\title[Study of the Ashkin Teller model with spins $S$ = $1$ and $\sigma$ = $3/2$ subjected to different crystal fields]%
{Study of the Ashkin Teller model with spins $S$ = $1$ and $\sigma$ = $3/2$ subjected to different crystal fields using the Monte-Carlo method
}
\author[Z. Amimer \textsl{et al.}]{Z. Amimer\refaddr{label1}, 
        S. Bekhechi\refaddr{label2}, B.N. Brahmi\refaddr{label2}, R. Boudefla\refaddr{label2}, H. Ez-Zahraouy\refaddr{label3}, 
        A. Rachadi\refaddr{label3}}
\addresses{
\addr{label1} Laboratory  of  Automatics, Faculty of Sciences, University of Tlemcen, BP119, Algeria
\addr{label2} Laboratory of Theoretical Physics, Faculty of Sciences, University of Tlemcen, BP119, Algeria
\addr{label3} Laboratory of Condensed Matter and Interdisciplinary Sciences (LaMCScl), Faculty of Sciences, Mohammed V University of Rabat, P.O. Box 1014, Morocco
}
\date{Received March  3, 2020, in final form June 18, 2020}

\begin{document}

\maketitle

\begin{abstract}
Using the Monte-Carlo method, we study the magnetic properties of the Ashkin-Teller model (ATM) under the effect of the crystal field with spins $S = 1$ and $\sigma = 3/2$. First, we determine the most stable phases in the phase diagrams at temperature $T = 0$ using exact calculations. For higher temperatures, we  use the Monte-Carlo simulation. We have found rich phase diagrams with the ordered phases: a Baxter $3/2$ and a Baxter $1/2$ phases in addition to a $\left\langle \sigma S\right\rangle$ phase that does not show up either in  ATM spin 1 or in ATM spin $3/2$ and, lastly, a $\left\langle \sigma\right\rangle  = 1/2$ phase with first and second order transitions.
\end{abstract}

\section{Introduction}

In recent years, the functioning of spins in different network structures has been a magnetic manifestation. It also allowed one to verify the nature of the phase transition as well as the critical behavior in the field of statistical mechanics \cite{A1}. In addition, the properties of magnetic materials and their technological applications such as thermomagnetic recording media and micro-electromechanical systems~\cite{2}are characterized by the phenomenon of mixed spins, which are well defined in the Ising model approach~\cite{3}. Studies of magnetic materials of mixed spins have been extended to the Ising model in the presence of a crystalline field and, specifically, are applied to the mixed spin ($1,3/2$). The latter studies have shown some interesting behaviors using an effective field theory. The results of the field theory study have shown that the mixed spin system has first order transition lines as well as offers tricritical and triple points. They also found out that the system  is of the  types \cite{4}. However, in the context of mixed spin ($1,3/2$), the Blume-Capel Ising model was realized when a first order transition line was found separating two ferromagnetic regions on a square cubic lattice \cite{5}. Using Monte Carlo simulations, it was shown that the interactions between the nearest neighbors of the Ising model $J_1$, $J_2$ and $J_3$ with frustrations are the main barriers to the transmission change in the increase in temperature and also indicate an Ashkin-Teller behavior. This study estimates the transition points at the critical point of Potts and confirms the first order transition behavior in the stabilization state of $J_3$ antiferromagnetic \cite{6}. In addition, they  conducted studies on the nature of the four-stage thermal phase transition degraded in a Monte Carlo simulation and the finite size scaling.  On the other hand, first-order behaviors are noted under Potts' critical points with four states, and thus, his work indicates that the four-state transition in the Ising antimagnetic model represents a similar transition \cite{7}. In this context and to properly describe the notion of phase transitions, Ashkin and Teller \cite{8} developed a very interesting model in these Ising systems and, thus, simplified the study of statistical mechanics. In this model, one could introduce the cooperative phenomena of quaternary alloys into a network, which is described by a Hamiltonian in a form suitable for magnetic systems  \cite{9}. Kramers and Wannier observed critical points of a particular case of the Ashkin-Teller model in which three of the four components are degraded \cite{10new}. Their hypotheses extended to the Ashkin-Teller model shown by Fan \cite{11new}, and tended to be that of Wegner who generally proved that the argument did not exist at a critical point. Hence, it is interesting to study closely the problem of transition in this model \cite{11}. In addition to this, Wagner proved that the Ashkin-Teller model was the equivalent of the alternating eight vertex model, which has not been solved exactly; only one critical line in the phase diagram of the isotropic Ashkin-Teller model is as precise as possible thanks to the duality relation found by Fan  \cite{11new}. One of the most interesting critical properties of this model is the non-universality of critical behaviour on the self-doubling lines in which the critical exponents evolve continuously  \cite{12}. The model is a two-dimensional system in which two layers of Ising spins interact with each other by a four-spin system. Within each of the models, there is an interaction at two turns between the nearest neighbors, i.e., a coupling of two Ising models that are characterized by the spins located in each cubic network site with a four-spin interaction parameter  \cite{13}. In particular, the numerical study of the Ashkin-Teller spin-1 model under the crystal field effect was carried out by Badehdah et al. \cite{14}. In addition, Wu and Lin found out a diversity of Ising type phase transitions of the anisotropic Ashkin-Teller model \cite{15}, and in this system, Bekhechi et al. analyzed the critical behavior of the Ashkin-Teller model using the averaged field and the Monte Carlo methods, with which they established that the $\left\langle \sigma\right\rangle $ phase appears in the isotropic case when the interactions are antiferromagnetic \cite{13}. 
Recently, the development of the Ashkin-Teller model has been achieved by the absorption of the selenium compound on the Ni surface \cite{16}, and the phase diagram obtained from the elastic DNA response \cite{17}. This model  also needs the study of thermodynamic properties of superconducting capsules (CuO$_{2}$ chips) \cite{18}.  Furthermore, the oxygen estimate can also be adapted in YBa$_{2}$Cu$_{3}$O$_{z}$ to the two-dimensional Ashkin-Teller model \cite{12}. This model has also been applied over the years in other fields such as chemical reactions in metal alloys~\cite{19}. In addition and because of the similarities to this model, which presents a complex and important phase diagram, different theoretical and numerical methods have been applied including Monte Carlo simulation \cite{20}, mean field approximation \cite{21}, effective field theory \cite{22}, matrix transfer method \cite{23} and renormalization group theory \cite{24}. This model can also describe phenomena \cite{14}.
In this paper, we essentially study the isotropic spin ($S=1$, $\sigma=3/2$) of the Ashkin-Teller model in the fundamental state at zero temperature ($T=0$) using the Monte Carlo method. We  have also examined this model at high temperatures using Monte Carlo simulations. We also check the stable phases of this model for different parameters $K_4$, $D_1$ and $D_2$. The paper is structured as follows; after an introduction, we have determined the fundamental state of the model and their basic diagrams. The subsequent section describes the Monte Carlo simulation and its formalism with the presentation of the phase diagram of the model. Then, we  discuss the results of the simulations. Finally, we  present a concluding section.

\section{Model and phase diagram of the fundamental state}

In this work, we consider the Ashkin-Teller model in the case of mixed spins $\sigma=3/2$ and $S=1$. We analyze this case under the effect of different crystal fields. Thus, this model is described by the following Hamiltonian:
\begin{equation}
H=-K_2\sum_{(\left\langle i,j\right\rangle)}(\sigma_i\sigma_j+S_i S_j )-K_4\sum_{(\left\langle i,j\right\rangle)}\sigma_i\sigma_j S_i S_j-D_1\sum_{i}S_i^2-D_2\sum_i\sigma_i^2\,,
\label{eq2.1}
\end{equation}
where the variables $\sigma_i$ and $S_i$ take the values ($\pm3/2$, $\pm1/2$) and ($\pm1,0$), respectively, and are located on the sites of a cubic lattice, $\left\langle i,j\right\rangle$ refers to a pair of nearest neighbor spins.
The first term of equation~(\ref{eq2.1}) refers to the bilinear interactions between the spins located at the sites $i$ and $j$ using the coupling parameter $K_2$. Moreover, the second term refers to the interaction of the four spins with the coupling constant $K_4$. The last term refers to the existence of two ionic crystal fields $D_1$ and $D_2$.
From the contribution of a pair of $S_1$, $S_2$, $\sigma_1$ and $\sigma_2$, the Hamiltonian is expressed as a sum of contributions of the nearest neighbors, we obtain the pair energy as follows:
\begin{equation}
E_P=-K_2 \left[ (\sigma_1\sigma_2)+S_1 S_2 \right] -K_4 (\sigma_1\sigma_2 S_1 S_2)-\frac{D_1}{2}(S_1^2+S_2^2)-\frac{D_2}{2} (\sigma_1^2+\sigma_2^2 ).
\end{equation}

According to the values containing the variables $S_i$ and $\sigma_i$, we extract 144 $(3^2 \times 4^2)$ possible configurations for the ground state at $T=0$. Using symmetry configurations, this number reduces to 24 configurations. For each set of parameters: $K_2$, $K_4$, $D_1$ and $D_2$, we select the configuration with minimal energy $E_p$. This leads to the phase diagram in the fundamental state ($T=0$). Different phases will be given in the form ($S_1,\sigma_1,S_2,\sigma_2$).
In what follows, we consider different situations by fixing one parameter and varying the others (the latters will be normalized by $K_2$).
In figure~\ref{fig-smp1}, we plot the phase diagram by varying the parameter $K_4/K_2$ as a function of $D_2/K_2$ (letting $D_1=0$): 
\begin{figure}[!t]
\centerline{\includegraphics[width=0.40\textwidth]{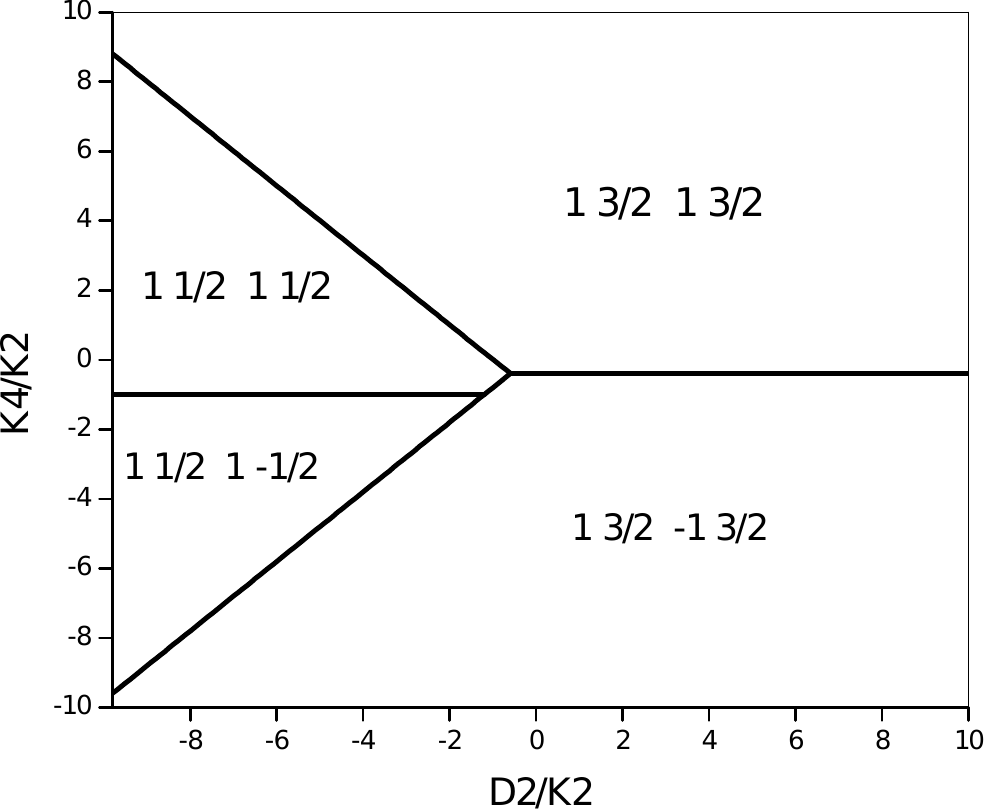} }
\caption{Phase diagram of the fundamental state in the case of $D_1=0$.} \label{fig-smp1}
\end{figure}

\begin{figure}[!b]
\centerline{\includegraphics[width=0.6\textwidth]{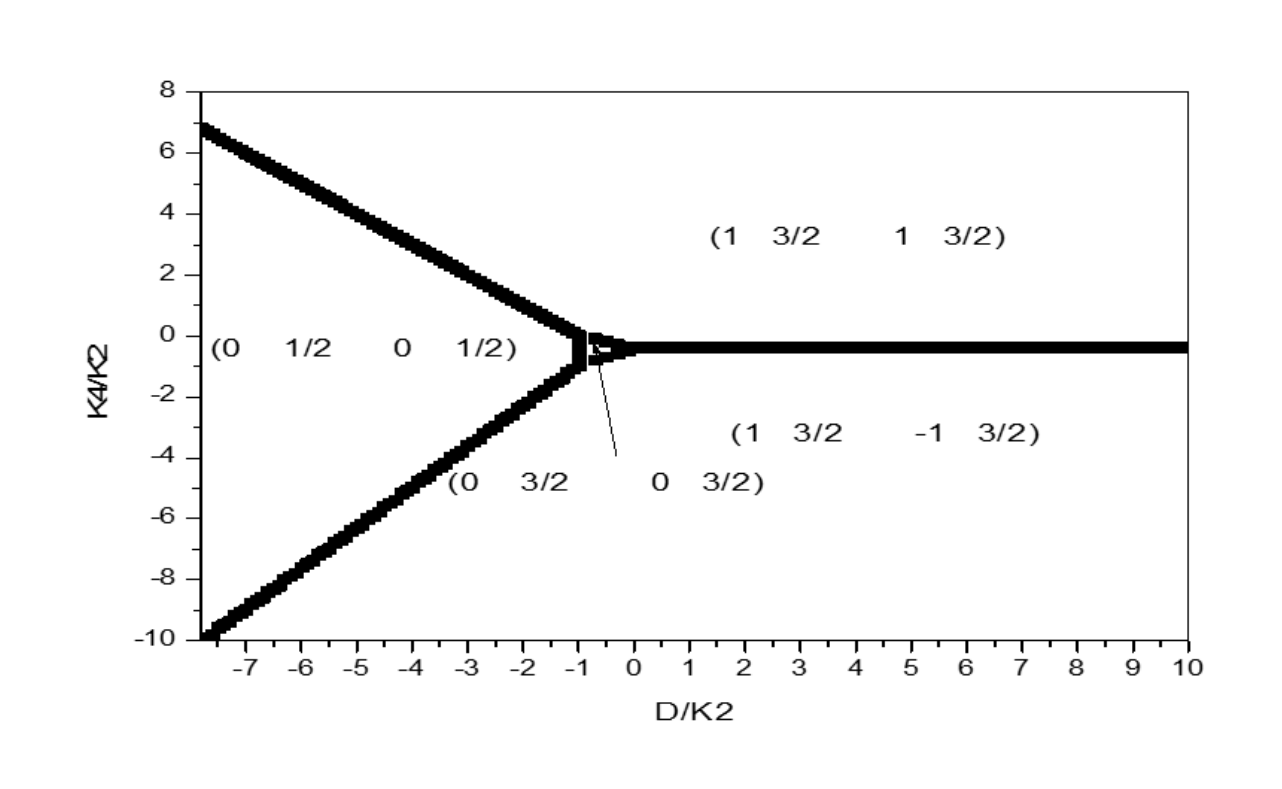}}
\caption{Phase diagram of the fundamental state in the case $D_1 = D_2 = D$.} \label{fig-smp3}
\end{figure}

\begin{itemize}
		\item 	For $D_2/K_2 < -1$, if $K_4/K_2 <-1$ and $K_4/K_2 > D_2/K_2 +1/4$: we observe that the $S_i$ spins are parallel such that $\left\langle S_i\right\rangle =1$ and $\sigma_i$ spins are antiparallel. Consequently, the stable phase obtained is the antiferromagnetic phase. For $K_4/K_2 >-1$ and $K_4/K_2 >-D_2/K_2 -1$. We can distinguish that the spins $S_i $ and $\sigma_i$ are both aligned in the same direction, so $\left\langle \sigma S\right\rangle =1/2$; this corresponds to the ferromagnetic phase.
	  \item In the case: $D_2/K_2 >-1$, if $K_4/K_2 <-4/9$, we can observe that the $\sigma_i$ spins are parallel such that $\left\langle \sigma\right\rangle = 3/2$ and $S_i$ spins are antiparalleled. Consequently, the stable phase obtained is the antiferromagnetic phase. Otherwise if $K_4/K_2 >-1$   and $K_4/K_2 >-D_2/K_2 -1$. We can distinguish that the spins $S_i$ and $\sigma_i$ are both aligned in the same direction, including $\left\langle S\right\rangle =1$ and $\left\langle \sigma\right\rangle =3/2$ so $\left\langle \sigma S\right\rangle =3/2$ while we have the ferromagnetic phase $3/2$.
\end{itemize}

\begin{figure}[!t]
\centerline{\includegraphics[width=0.40\textwidth]{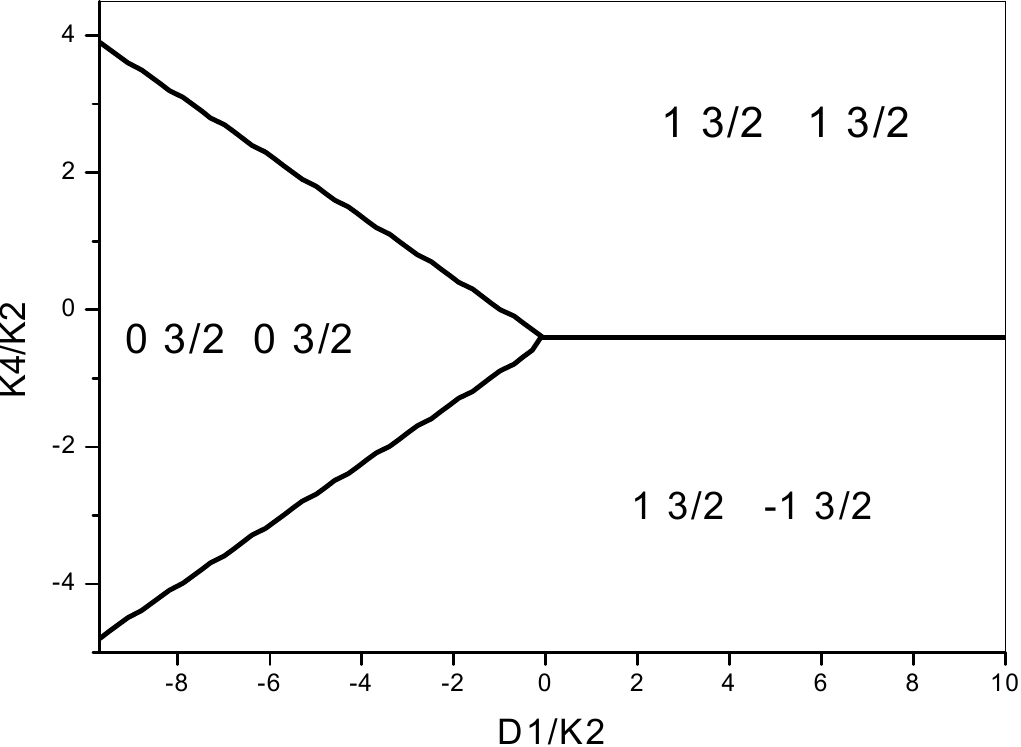}}
\caption{Phase diagram of the fundamental state in the case of $D_2=0$.} \label{fig-smp2}
\end{figure}

In the second situation, we obtain the figure which represents the variation of $K_4/K_2$ as a function of $D_1/K_2$. Lastly, we put $D_1= D_2=D$ and draw the diagram $K_4/K_2$ as a function of $D/K_2$ (figure~\ref{fig-smp3}).
Figure~\ref{fig-smp2} shows a phase diagram in the fundamental state and parameter variation $K4$ as a function of crystalline field D1 ($D2=0$).

This figure shows a phase diagram in the fundamental state and parameter variation $K_4$ as a function of crystalline field $D_1$ ($D_2=0$):
	\begin{itemize}
		\item 	For $D_1/K_2 <-0.1$ or $K_4/K_2 >4/9  D_1/K_2 -4/9$ and $K_4/K_2 >-4/9  D_1/K_2 -4/9$: we have a stable phase called the phase $\left\langle \sigma\right\rangle $ because we have the spins $S_i$ being equal to zero such that $\left\langle S\right\rangle  =\left\langle \sigma S\right\rangle =0$ and $\left\langle \sigma\right\rangle =-3/2$.
	\item For $D_2/K_2 >-0.1$, if $K_4/K_2 >-4/9$ and $K_4/K_2 >-D_2/K_2 -1$: we observe that the $S_i$ spins are parallel such that $\left\langle S\right\rangle =1$ so that the $\sigma_i$ spins are antiparallel or $\left\langle \sigma\right\rangle = 3/2$, consequently the stable phase obtained is the antiferromagnetic phase. Otherwise if $K_4/K_2 >-1$  and $K_4/K_2 >-D_2/K_2 -1$. We can observe that the spins $S_i$ and $\sigma_i$ are both aligned in the same direction, including $\left\langle S\right\rangle=1$ and $\left\langle \sigma\right\rangle =3/2$ so $\left\langle \sigma S\right\rangle =3/2$ while we have the ferromagnetic phase 3/2 as the stable phase.
\end{itemize}
The third case was made using $D_1=D_2=D$ (figure~\ref{fig-smp3}). We draw the diagram $K_4$ according to crystalline field $D$ such that:
\begin{itemize}
		\item	For $K_4/K_2 >-12/9  D/K_2 -12/9$ and $K_4/K_2 >-0.4$; we have $\left\langle \sigma\right\rangle =3/2$, $\left\langle S\right\rangle=1$ and $\left\langle \sigma S\right\rangle =3/2$ so $\left\langle \sigma\right\rangle = \left\langle \sigma S\right\rangle $ such that the spins $\sigma_i$ and $S_i$ are both parallel, then we have the Baxter 3/2 phase called the ferromagnetic Baxter phase (the stable phase).
	  \item For $K_4/K_2 <4/9  D/K_2 -4/9$ and $K_4/K_2 <-0.4$; in this part of the diagram we see that the spins $\sigma_i$ are parallel so that the spins $S_i$ are antiparallel. This means that we have an antiferromagnetic Baxter phase, which is always the Baxter phase (3/2).
	  \item For $K_4/K_2 <-12/9  D/K_2 -12/9$ and $K_4/K_2 >12/9  D/K_2 +4/9$, in this region we have $\left\langle \sigma\right\rangle =1/2$ and $\left\langle S\right\rangle=0 = \left\langle \sigma S\right\rangle =0$, so the phase here is the phase symbolized by $\left\langle \sigma\right\rangle $, because the spins $S_i$ are equal to zero and the spins $\sigma_i$ while the parallel ones designate phase 1/2.
	   \item For the zone that is noticed in the phase diagram and which is specificed by the equations: $K_4/K_2 >4/9  D/K_2 -4/9$ and $K_4/K_2 <-4/9  D/K_2 -4/9$ and if $D=-1$, we have the spins $\left\langle S\right\rangle=0$ and $\left\langle \sigma\right\rangle =3/2$, such that $S_i$ are equal to zero and $\sigma_i$ are parallel. Finally, we obtain the phase $\left\langle \sigma\right\rangle $ is the phase $\left\langle \sigma\right\rangle = 3/2$  which does not exist either in the case of mixed spin $-1/2$ \cite{13} or in the Ashkin-Teller model for spin-3/2 \cite{25}.
\end{itemize}

\section{The Monte Carlo simulation}
In our work, to determine the magnetic properties of the Ashkin-Teller model for non-zero temperatures, we use Monte Carlo simulations implemented with the Metropolis algorithm with periodic boundary conditions to update the lattice configurations.
We consider a 2d square lattice of $L\times L$ size which contains $N=L^2$ sites. We  performed the simulations for system size $L=30$. We performed simulations for certain values of the parameters $K_4$, $D_1$ and $D_2$ using $P=100000$ Monte Carlo steps after discarding the first $20000$ MCS for thermal equilibrium. The magnetization of the system is given by the formula:
\begin{equation}
|M_\alpha|=\left\langle |M_\alpha|\right\rangle =\frac{1}{N_P}\sum_c\sum_i\alpha_i (c)
\end{equation}
with $\alpha=\sigma,S,\sigma S$,
where $i$ runs over the lattice sites and $c$ runs over the obtained system configurations obtained. The lattice is updated by a sweep of the $N$ spins (the Monte Carlo step) after the system reaches thermal equilibrium.
The magnetic susceptibility relationship is given by:
\begin{equation}
\chi_\alpha=N/(K_\text{B} T) \left( \left\langle M_\alpha^2 \right\rangle -\left\langle |M_\alpha\right\rangle ^2 \right) 
\end{equation}
with $\alpha=\sigma,S,\sigma S$
and the Binder cumulant is given by:
\begin{equation}
U_\alpha=1-\frac{\left\langle M_\alpha^4 \right\rangle }{3\left\langle M_\alpha^2 \right\rangle ^2 }\,.
\label{eq3.3}
\end{equation}
Errors are deducted from the blocking method.

\section{Results and discussions}
\begin{figure}[!t]
\centerline{\includegraphics[width=0.90\textwidth]{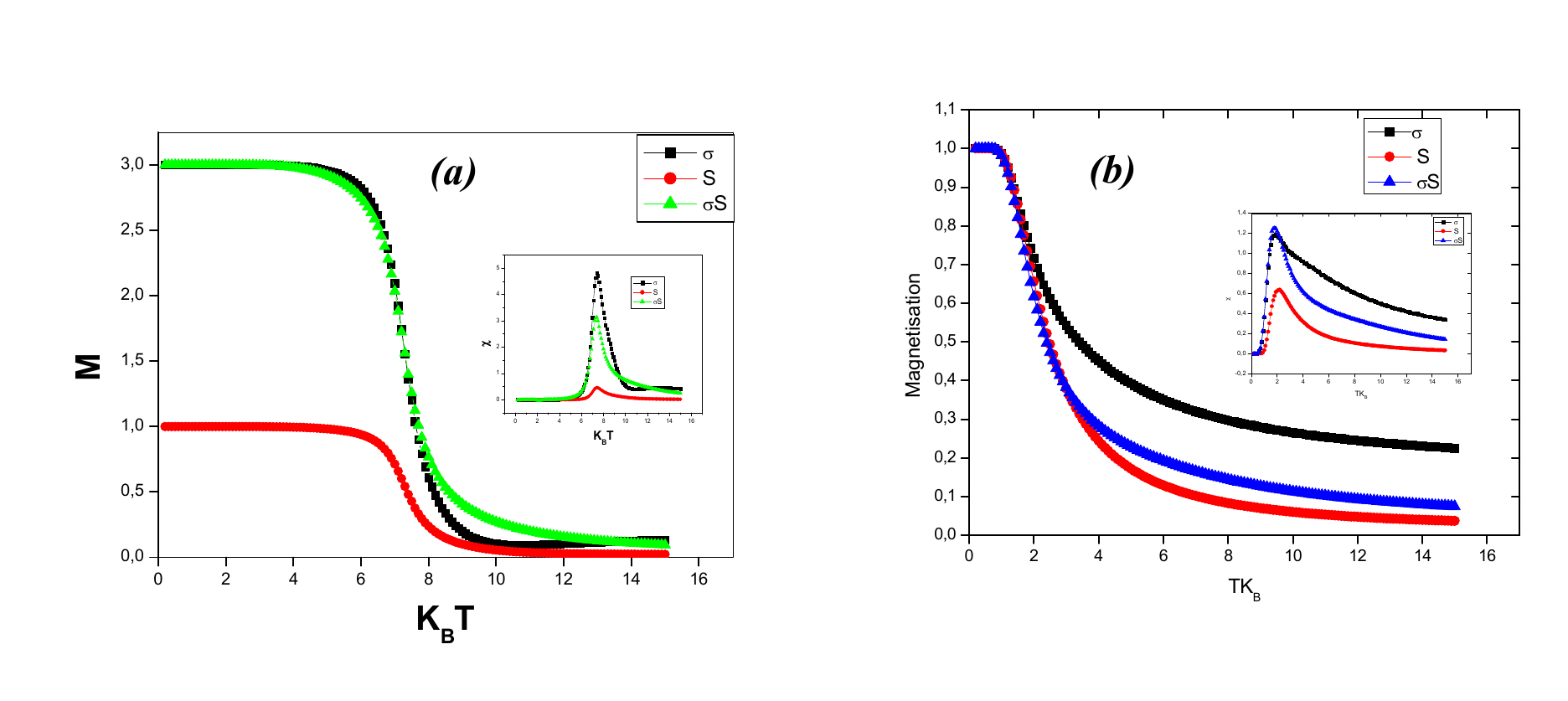}}
\caption{(Colour online) Phase diagram shows the magnetization (the parameters $\sigma, \,\,S,\,\, \sigma S$) as a function of the temperature, with the size $L = 30$, with crystal field $D_1=0$; we take $K_4 = 1$ and for a) $D_2 =6$ and b) $D_2=-6$. Continuous behavior is observed signaling second order transition.} \label{fig-smp4}
\end{figure}

\begin{figure}[!t]
\centerline{\includegraphics[width=0.90\textwidth]{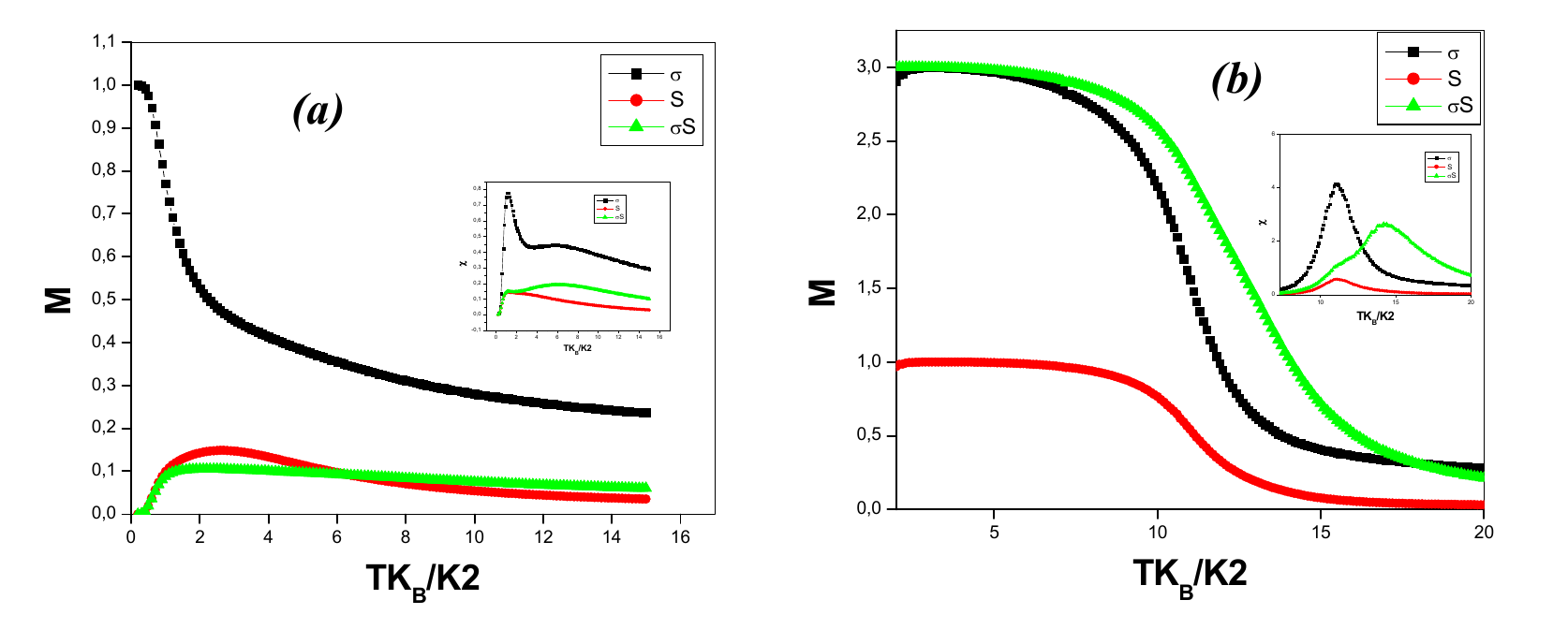} }
\caption{(Colour online) The magnetization (for the parameters $\sigma,\,\, S,\,\, \sigma S$) as a function of the temperature, with system size $L = 30$, with crystal field $D_1 = D_2 = D$, in the two cases a) $K_4 = 1$, $D = -6$ and b) $K_4 = 3$, $D = -2$ a partially ordered is observed at high temperature.} \label{fig-smp5}
\end{figure}

We obtain the magnetization behavior as a function of temperature as well as the susceptibilities of the studied system for different values of the coupling parameters.
As shown in figure~\ref{fig-smp4} our MC results  at low temperature show a ferromagnetic Baxter phase $(S_1 \sigma_1 S_2 \sigma_2)$=(1 3/2 1 3/2) with $\left\langle \sigma S\right\rangle =3/2$ (figure~\ref{fig-smp4}a) and a ferromagnetic Baxter phase (1 1/2 1 1/2)  with $\left\langle \sigma S\right\rangle =1/2$ (figure~\ref{fig-smp4}b) as expected from the $T=0$ phase diagram (figure~\ref{fig-smp1}), where we find a new partially ordered phase $\left\langle \sigma S\right\rangle $ identified by $\left\langle \sigma\right\rangle =\left\langle S\right\rangle=0$, and $\left\langle \sigma S\right\rangle \ne 0$. For high temperatures in both cases, the system becomes disordered. The critical transition temperature is estimated from the maximum of the susceptibility associated with the different magnetization. We found for the case (a) $T_c=7.39$ and for case (b) $T_c=1.89$. In addition, the transition between the phases mentioned is always of second order due to the continuity of the order parameters across the transition line.
In figure~\ref{fig-smp5}, the first case (a) at low temperature, we have $\left\langle \sigma\right\rangle $ = 1/2, $\left\langle S\right\rangle=\left\langle \sigma S\right\rangle  = 0$ corresponds to the phase (0 1/2 0 1/2). However, at high temperature the system undergoes a transition to the paramagnetic phase. For the second case (b) $D=-2$ and $K_4= 3$, the ground state is the ferromagnetic Baxter phase (1 3/2 1 3/2). The susceptibility plot shows a peak corresponding to $\left\langle \sigma\right\rangle $ and $\left\langle S\right\rangle$ at the critical temperature $T_{c1}=11.09$; by contrast, the susceptibility corresponding to $\left\langle \sigma S\right\rangle $ shows a distinct peak at the transition temperature $T_{c2}=14.19$, clearly defining a partially ordered phase $\left\langle \sigma S\right\rangle $ at high temperature separating the disordered phase from the Baxter phase.

\begin{figure}[!t]
\centerline{\includegraphics[width=0.50\textwidth]{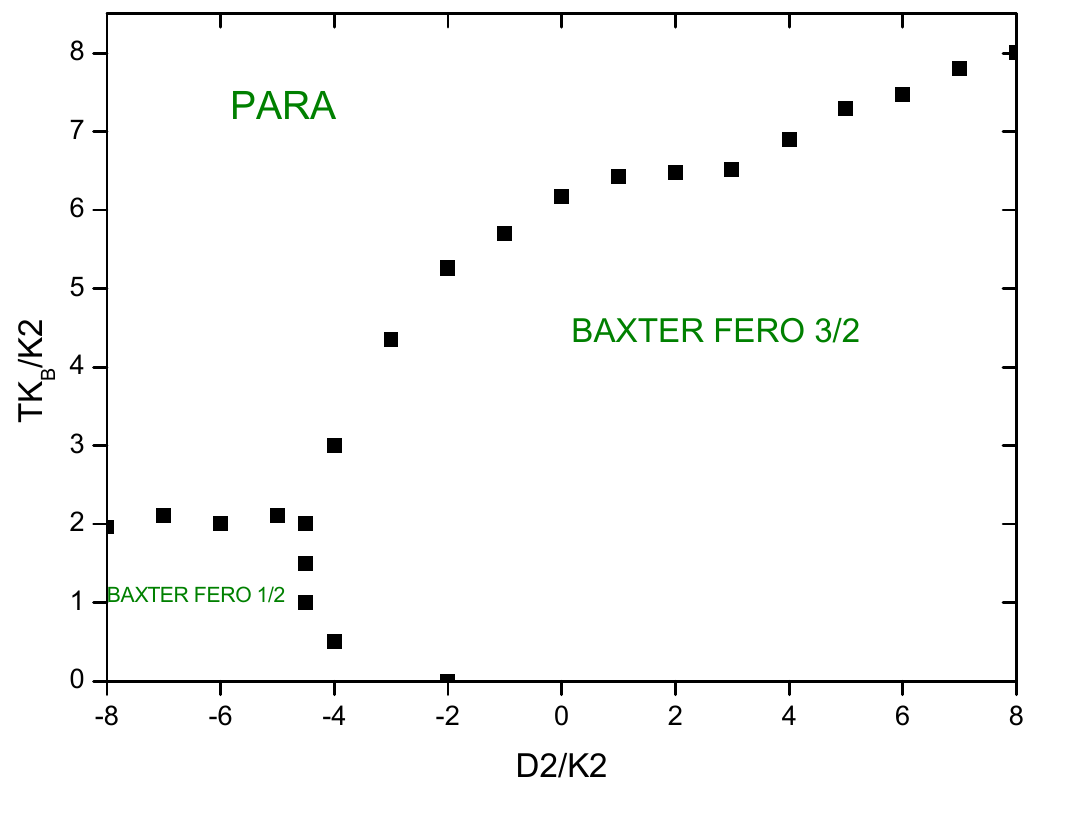} }
\caption{(Colour online) Phase diagram in the plane $(D_2 / K_2, T / K_2)$ for $K_4 = 1$ by MC simulation with $L = 30$.} \label{fig-smp6}
\end{figure}

\begin{figure}[!t]
\centerline{\includegraphics[width=0.55\textwidth]{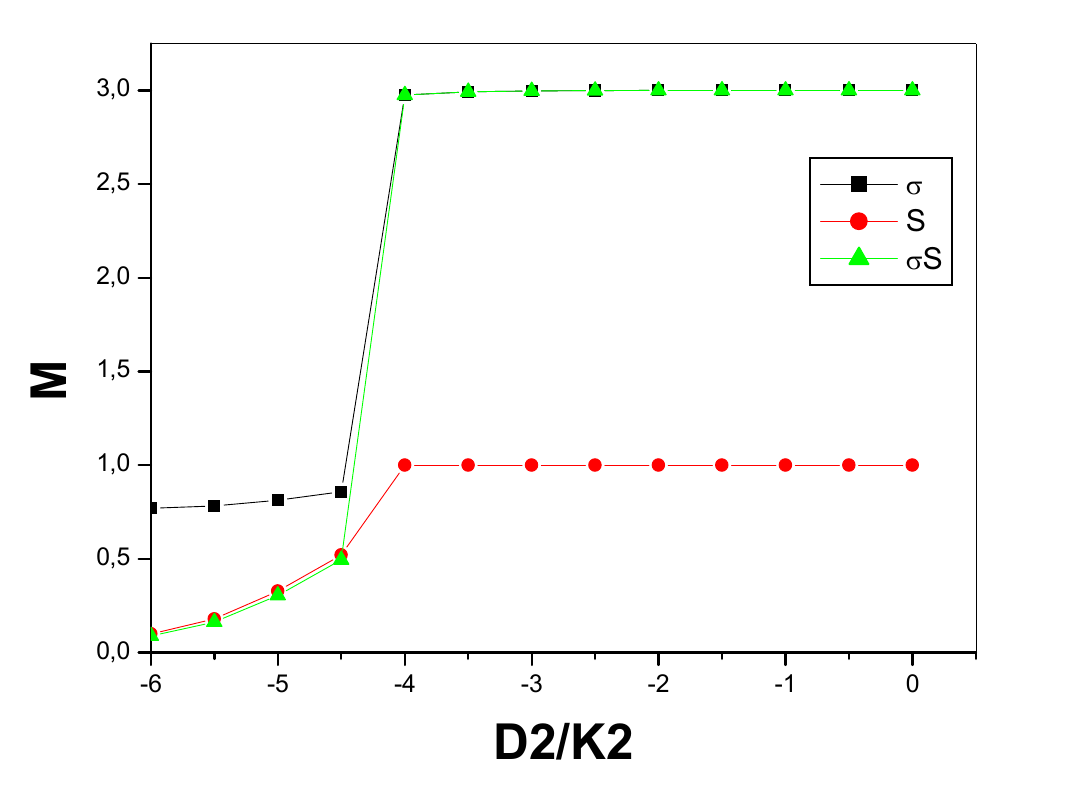} }
\caption{(Colour online) The magnetization (the parameters $\sigma$, S, $\sigma S$) as a function of the temperature with system size $L = 30$ with $T=1$.} \label{fig-smp8}
\end{figure}

The phase diagram in figure~\ref{fig-smp6} shows the stable phases at different temperature in the plane $(D_2/K_2,\\T/K_2)$ in the case $D_1=0$ for the coupling parameter $K_4=1$, we found that for low values of $D_2/K_2$, two phases are separated by a first order transition line. This is shown in figure~\ref{fig-smp8}. The verification of the phase transition nature is determined from the discontinuity or continuity of the order parameters \cite{24}.The two phases are: ferromagnetic Baxter $1/2$ and ferromagnetic Baxter $3/2$. The former phase $(\sigma=1/2)$ was neither found for this ATM model with spin-1/2 \cite{9} nor in the Ashkin-Teller model with spin-3/2 \cite{25}. At high temperature, a second order transition to the paramagnetic disorder phase takes place. The Baxter ferromagnetic-3/2 is stable for large values of $D_2/K_2$.
\begin{figure}[!t]
\centerline{\includegraphics[width=0.48\textwidth]{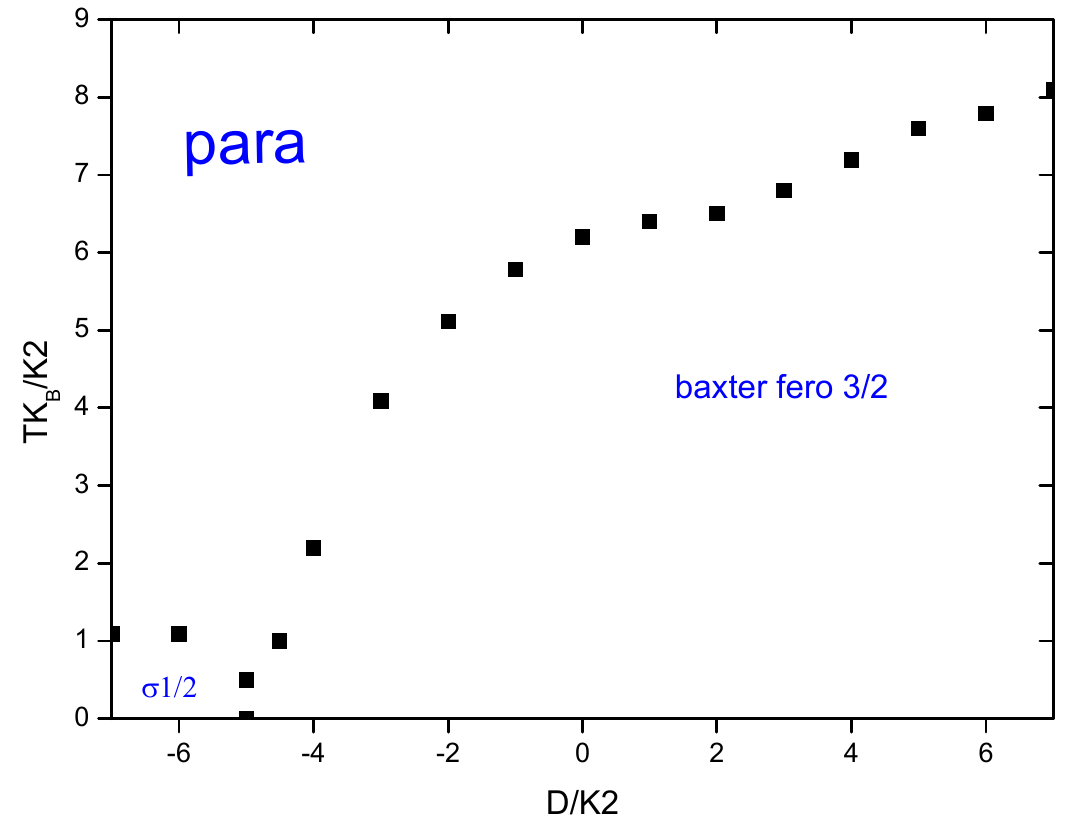} }
\caption{(Colour online) Phase diagram in the plane $(D / K_2, T / K_2)$ for $K_4 = 1$ by MC simulation with $L = 30$.} \label{fig-smp7}
\end{figure}

\begin{figure}[!t]
\centerline{\includegraphics[width=0.50\textwidth]{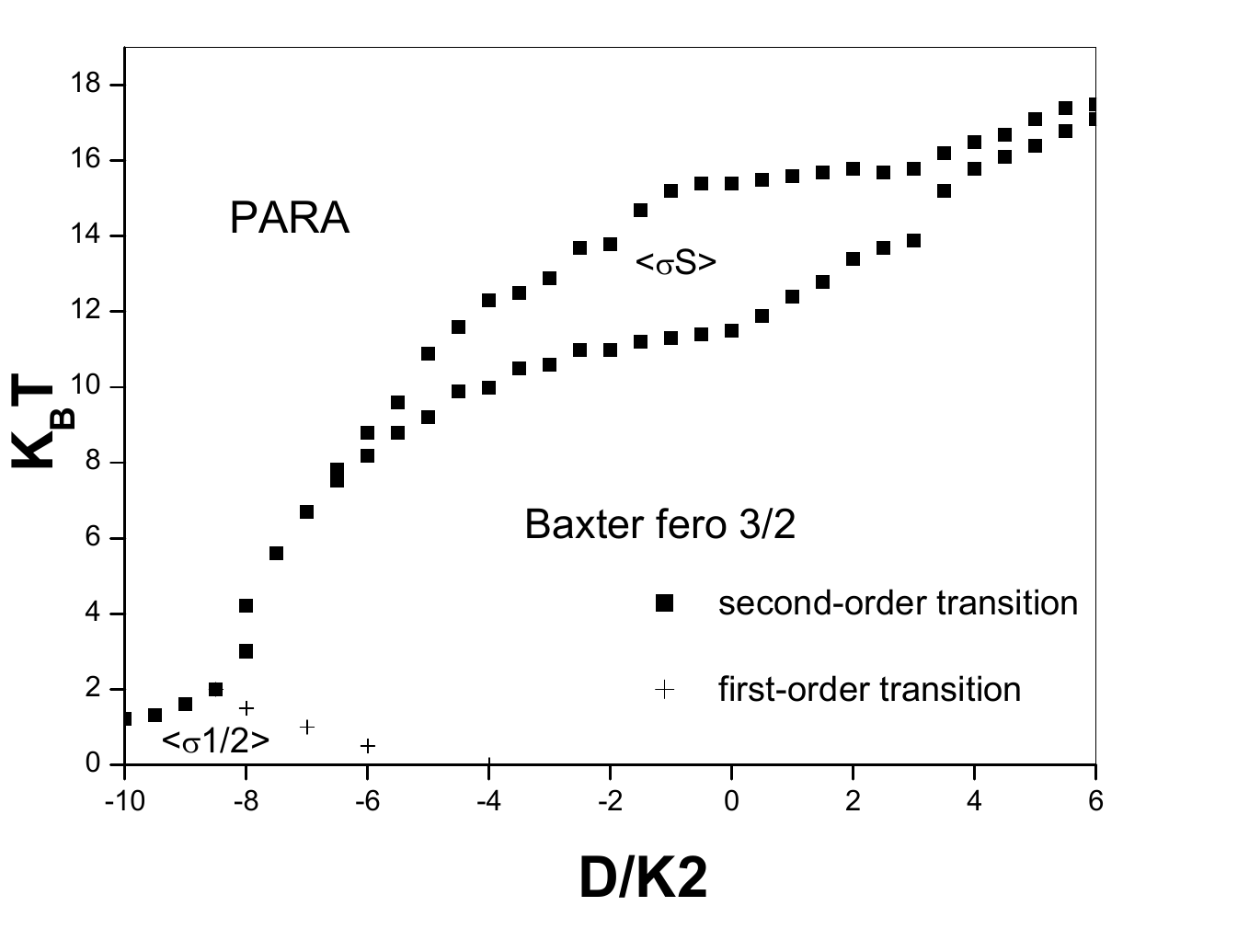} }
\caption{In-plane phase diagram $(D / K_2, T/ K_2)$ for $K_4 = 3$ by MC simulation with $L = 30$.} \label{fig-smp9}
\end{figure}

In figure~\ref{fig-smp7} we plot the phase diagram in the $(T/K_2,D/K_2)$ plane. We found out a similar form of the phase diagram as in figure~\ref{fig-smp6} except that the low temperature low $D$ phase is now $\left\langle \sigma\right\rangle =1/2$ phase with a first transition line to the ferromagnetic Baxter $3/2$ for large values of $D$. We also note  that the $\sigma-1/2$ phase was not found in the Ashkin-Teller for spin-3/2 \cite{25}.
Moreover, when the coupling parameter values are increased $K_4/K_2=3$ for the case of $D_1=D_2=D$ with the growth of the values of $D/K_2$. In figure~\ref{fig-smp9} we found out a partially ordered $\left\langle \sigma S\right\rangle $ phase, figure~\ref{fig-smp5}b, between the ordered phase of Baxter ferromagnetic 3/2 and the disordered paramagnetic phase at high temperature. These phases were illustrated in previous cases(figure~\ref{fig-smp6}, \ref{fig-smp7}). This new phase $\left\langle \sigma S\right\rangle $ was also found in the model Ashkin-Teller mixed of spin-1/2 \cite{13} but not in the Ashkin-Teller for spin-3/2 \cite{25}. By decreasing the crystal field $D/K_2$, we found a transition from Baxter ferromagnetic $3/2$ to the paramagnetic phase which is of first order type. We also observe  the same phase $\left\langle \sigma \right\rangle=1/2$ as in figure~\ref{fig-smp7} which separates high temperature paramagnetic phase in low values of $D/K_2$ by a second order transition, and is separated with a ferromagnetic Baxter $3/2$ phase by a first order transition.

\begin{figure}[!t]
\centerline{\includegraphics[width=1.1\textwidth]{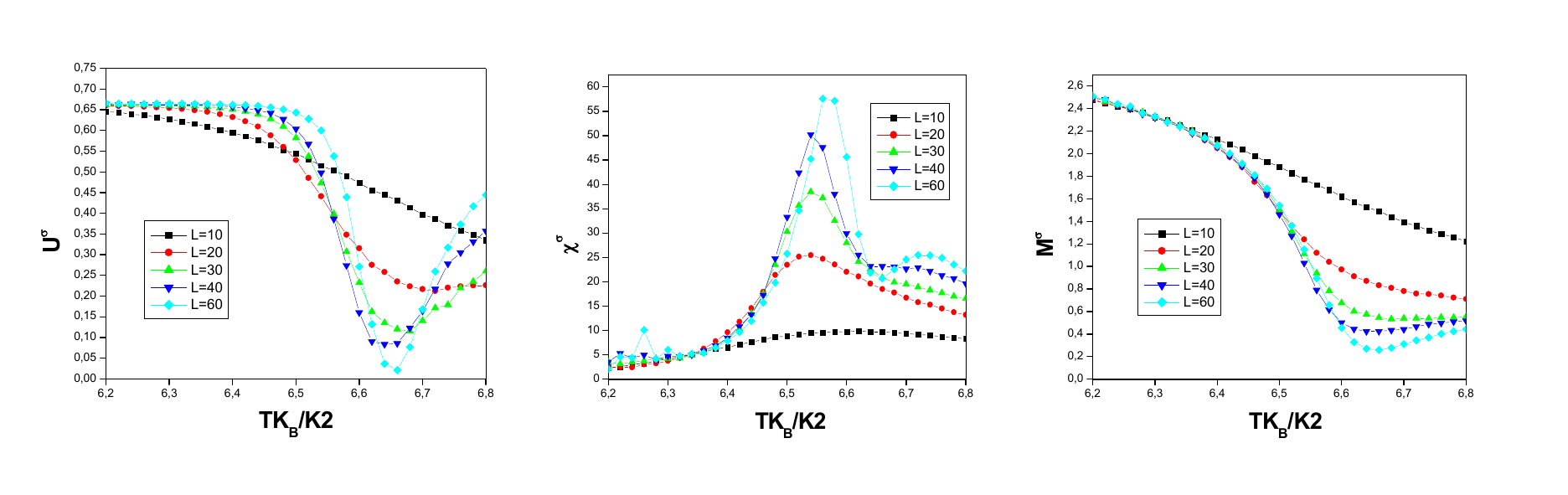}}
\caption{(Colour online) Dependance of the accumulate (a), the susceptibility (b) and the magnetization (c) for different choices of the size $L$ as a function of temperature in the case $K_4/K_2 = 1$ and $D_2/K_2=3$ $(D_1 = 0)$ for the  order parameter $\sigma$.} \label{fig-smp10}
\end{figure}
The phase transition points to a function of temperature at the fixed values of the coupling $K_4/K_2$ as well as the crystalline field $D_2/K_2$, which were pre-located from the points of intersection of the cumulative curves of Binder specified by the equation~(\ref{eq3.3}).
We show these cases for $D_2/K_2=3$ $(D_1/K_2=0)$ for different size of $L= (10,\, 20,\, 30,\, 40,\, 60)$ and the susceptibilities as a function of temperature as well as the Binder cumulant as a function of temperature figure~\ref{fig-smp10} plotted. It is noted in figure~\ref{fig-smp10} that the peaks when $L$ increases, show the transition point which means the critical temperature or the change of phase transition.
Nevertheless, the Binder accumulate curves whose figure~\ref{fig-smp10} shows that there is an intersection point that defines the critical temperature knowing that $T_c=6.55$.

\section{Conclusion}
In order to well describe  the magnetic properties of Ising typical systems in statistical mechanics, and within this framework, we analyzed the  Ashkin-Teller model with spins $(1, 3/2)$ on a cubic lattice under the effect of different crystalline fields and the coupling parameters which are defined in equation~(\ref{eq2.1}). The first step of our study  was the most important of the stable phase in the fundamental state (zero temperature) in three cases of crystalline field; the system undergoes a first-order phase transition between these stable phases because we noticed that we have phases that did not exist in ATM spin-1/2. On the other hand, when the temperature is non-zero, we have processed the AT model by the Monte Carlo simulation, specifically, using the Metropolis method. As a result, we  found out that the coupling parameter values were fixed and the crystal field varied with the temperature variation. We also found out that we  had the second-order phase diagram, which contained stable phases such as the Baxter phase 3/2 as well as the paramagnetic phase in  different cases of crystalline field in the parameter space ($K_4/K_2$, $D_1/K_2$, $D_2/K_2$, $D/K_2$, $T/K_2$) delimited by lines with multicritical points. Crucially, we found a new phase in the phase diagram in space ($K_4/K_2$, $D/K_2$, $T/K_2$). Finally, we verified the phase transition nature of this model, which is of second order phase transition of Ising type.

\ukrainianpart
\title
{Вивчення моделі Ашкіна-Теллера зі спінами  $S$ = $1$ і $\sigma$ = $3/2$ під дією різних критсалічних полів із застосуванням методу Монте Карло
}
\author{З. Амімер\refaddr{label1}, 
        С. Бехеші\refaddr{label2}, Б.Н. Брахмі\refaddr{label2}, Р. Будефла\refaddr{label2}, Г. Ез-Захрої \refaddr{label3}, 
        A.Рашаді\refaddr{label3}}
\addresses{
\addr{label1} Лабораторія автоматики, факультет природничих наук, Університет м. Тлемсен, BP119, Алжир
\addr{label2} Лабораторія теоретичної фізики, факультет природничих наук, Університет м. Тлемсен, BP119, Алжир
\addr{label3} Лабораторія конденсованої матерії і міждисциплінарних наук (LaMCScl), факультет природничих наук, Університет ім. Мохаммеда V, м. Рабат, Марокко 
}
\newpage
\makeukrtitle 

\begin{abstract}
\tolerance=3000%
Використовуючи метод Монте Карло, досліджено магнітні властивості моделі Ашкіна-Теллера (МАТ) зі спінами  $S$ = $1$ і $\sigma$ = $3/2$ під дією  кристалічного поля. Спочатку визначено найстійкіші фази на фазових діаграмах при температурі  $T = 0$ з допомогою точних обчислень. При вищих температурах ми використовуємо моделювання методом Монте Карло. Знайдено багаті фазові діаграми із впорядкованими фазами: фазу Бахтера 
 $3/2$ і фазу Бахтера  $1/2$ додатково до фази $\left\langle \sigma S\right\rangle$, яка не з'являється ні в МАТ зі спіном $1$, ні в МАТ зі спіном $3/2$ і, нарешті, фазу $\left\langle \sigma\right\rangle  = 1/2$ з переходами першого і другого роду.
\keywords модель Ізінга, Ашкін-Теллер, спін-$1$, спін-$3/2$, Монте Карло
\end{abstract}
\lastpage

\end{document}